\documentclass[twocolumn,11pt,cites,graphicx,floatfix,amsmath,amssymb]{article}
\usepackage{amssymb}
\usepackage{amsmath}
\usepackage{graphicx}
\usepackage{epstopdf}

\title{Highly Anisotropic Vorticity Aligned Structures in a Shear Thickening Attractive Colloidal System}
\author{Chinedum O. Osuji\\Department of Chemical Engineering\\Yale
University, New Haven CT 06511\\E-mail: chinedum.osuji@yale.edu \and
David A. Weitz \\School of Engineering and Applied Sciences\\
Harvard University, Cambridge MA 02138.
\\Email:weitz@seas.harvard.edu}
\begin{document}

\maketitle
\renewcommand{\thefootnote}{\fnsymbol{footnote}}

\noindent

Vorticity aligned cylindrical flocs of carbon black particles are
formed in steady flow at low shear rates, and strikingly, appear as
transient structures in the flow response of gels produced by the
quenching of high rate shear thickening flows.
\\

Structure formation in response to flow is common in soft materials
and complex fluids and is dependent on the nature of the cohesive
forces in the system, time scales for diffusive motions and
hydrodynamic interactions. For Brownian systems with hard sphere or
repulsive interactions, distortion of the microstructure due to an
imposed shear rate occurs when the timescale for flow is smaller
than that for diffusion. The relative strength of flow forces
compared to thermal or diffusive forces can be gauged by the
dimensionless P\'{e}clet number

\begin{equation}
Pe=\frac{\eta\dot\gamma a^{3}}{k_{B}T} \label{eq:Peclet_number}
\end{equation}

For $Pe \ll1$ the system displays isotropic arrangements of
particles, but for $Pe>1$ the timescale associated with flow
dominates that for diffusive relaxation. Under these conditions, the
system becomes anisotropic due to separation of particles along the
elongational axis and aggregation along the compressional axis
\cite{COO:Vermant2005}. String formation by alignment along the flow
direction is typical in dilute systems at high shear rates and
further structuring may take place by periodic arrangement of these
strings in the vorticity-gradient plane. In more concentrated
suspensions, hydrodynamic interactions lead to the formation of
hydro-clusters and a shear thickening or jamming transition
\cite{COO:MelroseBall2000,COO:ZukoskiChow1995_1}. In systems with
constitutive instabilities due to a negative slope in the
stress-strain curve, co-existence between different deformation
states under a single applied stress gives rise to shear banding
along the velocity gradient direction \cite{COO:Vermant2001}. For
attractive systems, structure formation as discussed above is also
found. Additionally, the alignment of particles or constituents
along the vorticity axis has been reported and has emerged as a
somewhat general phenomenon
\cite{COO:Pasquali_PRL2004,COO:Hobbie_PRE2004}. Shear thickening
though, is not generally observed in colloidal systems with
sufficiently strong attractive interactions to form flocculated gels
and is presumed not to occur \cite{COO:Zukoski2004_1}. It has been
reported in a handful of systems to date
\cite{potanin2004tar,kanai1995ntf,COO:Kato2001_1}, although the
complex composition of the suspensions studied in some cases makes
it difficult to unambiguously study the mechanism.

We have studied the steady state flow behavior of dilute, simple
hydrocarbon dispersions of carbon black particles and, surprisingly,
observe shear thickening above a composition dependent critical flow
rate, $\dot\gamma_{c} \approx 10^{2}-10^{3} s^{-1}$. The shear
modulus of gels formed by pre-shearing above the critical shear
thickening rate displays an interesting power law dependence on the
stress applied during the pre-shear. This is well accounted for by
the dependence of the cluster size on the applied shear stress, and
the resulting increase in cluster number density
\cite{COO:Osuji_PRE2007}. Subsequent deformation of these shear
thickened gels at low shear rates produces highly oriented vorticity
aligned flocs that gradually break down into small isotropic
clusters over timescales of $\approx$ 300 seconds.

We use tetradecane (Aldrich Chemical Co.) dispersions of 2 to 8
wt.\% of 0.5 $\mu m$ carbon black particles (Cabot Vulcan XC72R)
with a fractal dimension $d_{f}=2.2$. Rheological data reported here
were recorded using an AR-G2 (TA Instruments) rheometer in strain
control, although separate experiments were also run for
verification purposes on a pure-strain controlled ARES-LS1
instrument. A variety of geometries, standard and roughened, were
used to check for and to mitigate the impact of possible wall slip.
Although from our measurements it appears that wall slip is not
significant, its prevalence in other studies on ``sticky'' colloidal
gels makes the interpretation of data at low shear rates troublesome
\cite{COO:Khan_wall_slip_JRheol2003}. A more detailed study of the
low shear rate regime is underway. Optical observations of
microstructures under shear were made using a Bohlin CS rheometer
with a transparent base-plate through which images were recorded (by
reflection off the rheometer top tool) using a CCD camera.

\begin{figure}[h]
\begin{center}
\includegraphics[width=70mm, scale=1]{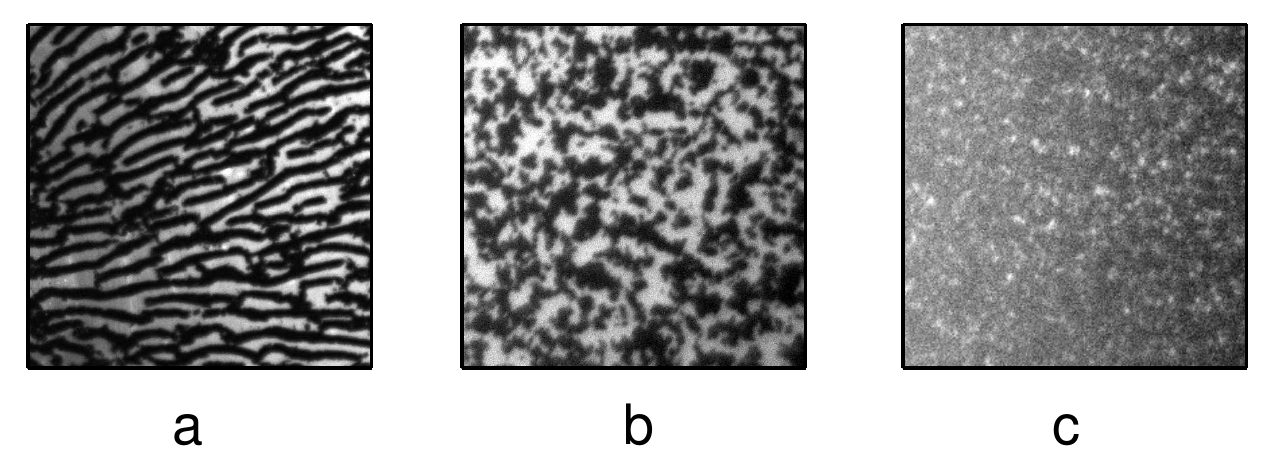}
\caption{Microstructure under shear in parallel plate geometry with
a gap size of 100 $\mu m$ for 3 wt.\% sample. (a) $\dot\gamma =
6.67s^{-1}$ 40 mm field of view showing cylindrical flocs aligned
along the vorticity direction. (b) 1.5 mm field of view, $\dot\gamma
= 133s^{-1}$ during shear (c) 1.5 mm field of view, $\dot\gamma =
1330s^{-1}$ during shear. To obtain sufficient light transmission
through the optically dense sample in (c), the illumination was set
several times higher than that used for (b). Binning of pixels on
the CCD was used to decrease the required exposure time and so the
resolution of image (c) is half that of (b).}
\label{optical_rheology}
\end{center}
\end{figure}

\begin{figure}[ht]
\begin{center}
\includegraphics[width=70mm, scale=1]{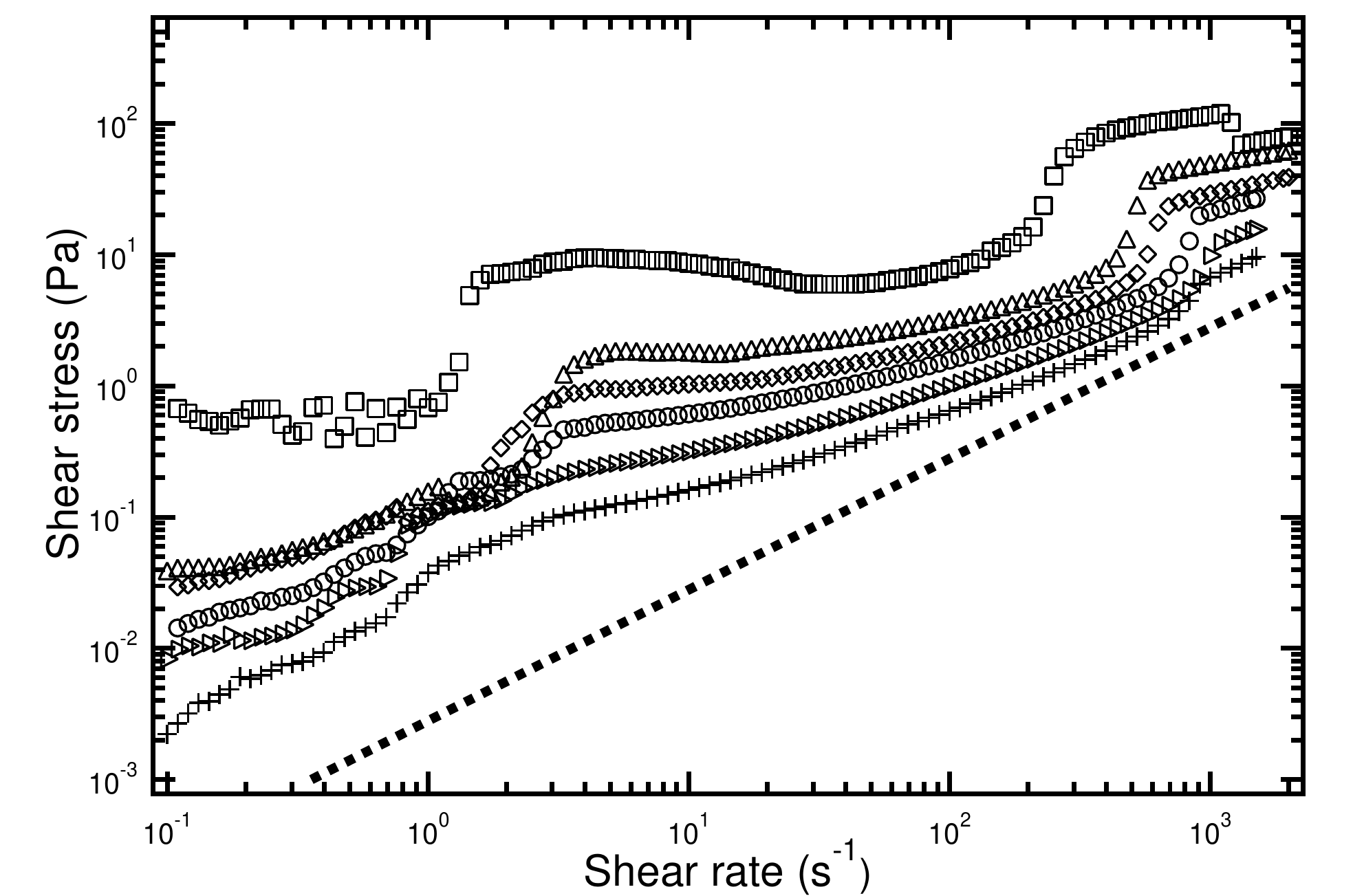}
\caption{Steady state flow curve of carbon black dispersions from 2
to 8 wt. \%.$+:2\%$, $ \triangleright:3\%$, $\circ:4\%$,
$\diamond:5\%$, $\triangle:6\%$, $\square:8\%$ The dotted-line shows
the stress due to the background viscosity of the solvent.}
\label{flow_curves}
\end{center}
\end{figure}

The flow curves for samples in steady shear display thixotropy, as
is typical for flocculating systems. Steady state flow curves of the
shear rate dependent stress display a composition-dependent,
continuous shear thickening transition at high shear rates,
$\dot\gamma_{c}\approx 10^{2}-10^{3}$, with distinct increases in
the stress and thus the viscosity of the dispersions with increasing
shear rate, Figure \ref{flow_curves}. From optical observations,
Figure \ref{optical_rheology}, at low shear rates $\dot\gamma
\lesssim 1s^{-1}$, the system is composed of large pieces of the
fractured gel which gradually break into smaller pieces with
increasing shear rate. Around $\dot\gamma \approx 10^0-10^1 s^{-1}$,
we observe a slight, if irregular increase in viscosity with
increasing shear rate. Coincident with flow in this region, we
observe the aggregation of clusters along the vorticity axis, and
the formation of elongated flocs which roll with the shear flow.
Such flocs are also observed in long observations overnight at
constant shear rate, indicating that, within the experimental time
frame, they represent the steady state response of the system to
flow in this range of shear rates. Continued increase of the shear
rate results eventually in dissolution of the cylindrical structures
into isotropic clusters. These clusters densify until
$\dot\gamma_{c}$ beyond which there is a transition to the shear
thickening flow as breakage of the aggregates is observed,
consistent with an increase in the effective volume fraction of
particles in the dispersion.

\begin{figure}[h]
\begin{center}
\includegraphics[width=70mm, scale=1]{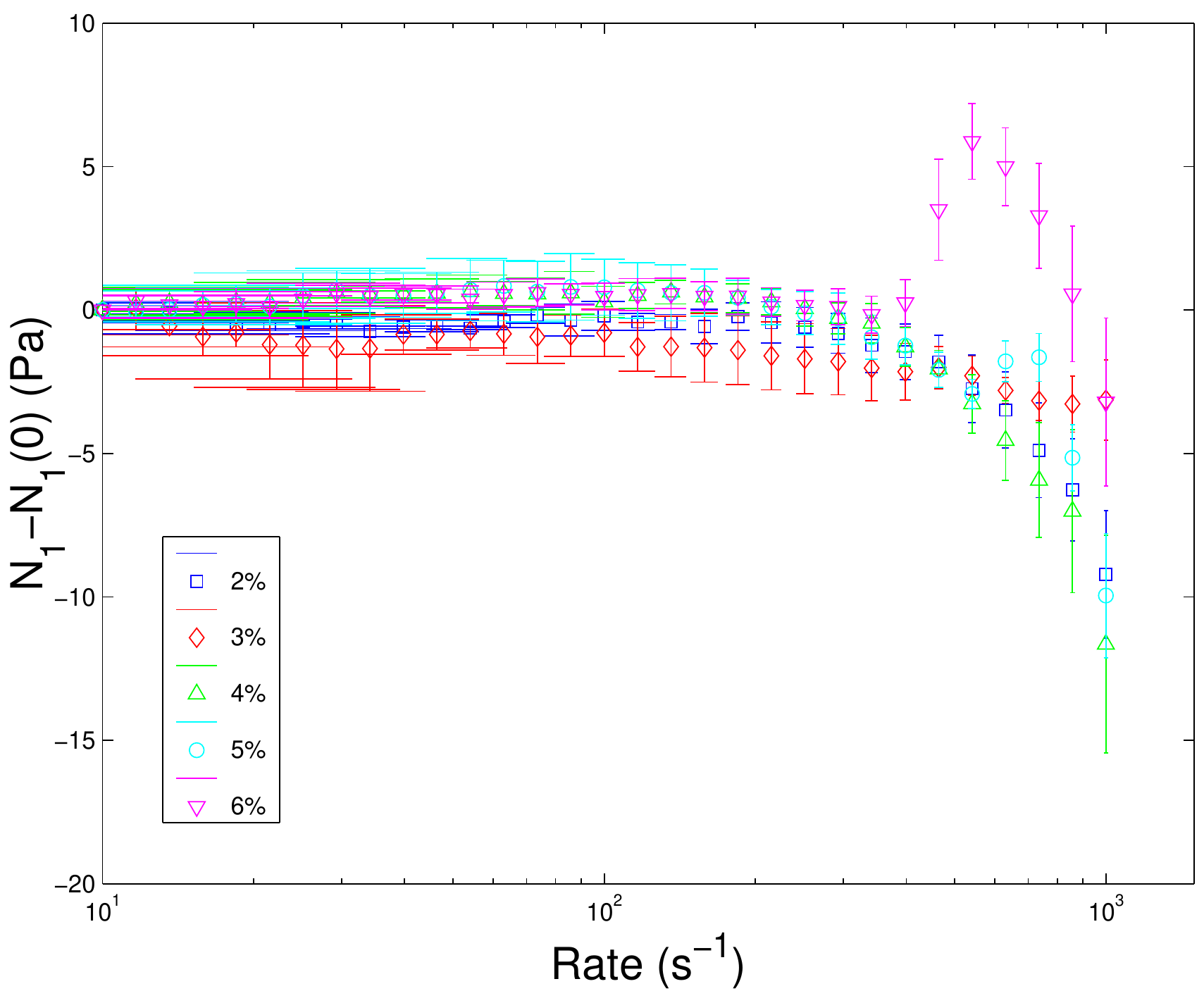}
\caption{First normal stress difference (relative to starting value,
$N_1(0)$)for various compositions as a function of shear rate. Data
are averaged over 4 runs. Error bars correspond to one standard
deviation.} \label{normal_stresses}
\end{center}
\end{figure}

Previous studies have implicated negative first normal stress
differences, $N_{1}=\sigma_{11}-\sigma_{22}$, in the formation of
vorticity aligned structures in flocculating systems
\cite{COO:Pasquali_PRL2004,COO:Hobbie_PRE2004}. Normal stresses
arise due to elasticity in the flowing suspension with elongational
stiffness and bending rigidity both contributing
\cite{COO:Dhont_PRE2006,COO:Migler_NatMat2004,COO:Shelley_PRL2001}.
In our system, we are unable to identify any systematic changes in
normal stress at low shear rates. It may simply be that the stresses
are too small to be accurately measured in our rheometer. In the
shear thickening regime however, samples develop large negative
normal stresses, on the order of $N_{1}\approx 2-10.\sigma_{shear}$,
depending on composition, due to the large inertia developed at
these high shear rates. Abrupt cessation of shear in this regime
produces a quench into a shear-thickened gel state where,
remarkably, both shear stresses and normal stresses were found to
persist for long times ($>$ 1 hour) under quiescent conditions and
to dissipate rapidly on application of shear flow. We have studied
the dynamics of the relaxation of the internal shear stresses and
find that they relax with a power law dependence $\sigma_i\sim
t^{-0.1}$ over timescales as long as $10^3-10^4$s
\cite{COO:Osuji_PRE2007}. Accurate measurement of the modest normal
stresses in dilute suspension rheology is made difficult by the lack
of better than $\approx$ 10-20 Pa resolution available on most
commercial rheometers, which rely on a strain gauge style mechanism
that is usually not stabilized against temperature fluctuation. The
presence of a large inertial contribution
\cite{COO:Porter_inertial_N1_RheolActa1977},
$N_{1}^{inertial}=-3/20\,\rho\omega^2R^2$ to our signal further
complicates this in our case, but from more concentrated samples,
$\phi>4$ wt.\%, we could reliably extract the sample contribution to
the normal stress, Figure \ref{normal_stresses}. Correction of the
data for inertial contributions shows that a positive normal stress
difference is developed at the onset of shear thickening, consistent
with the idea that shear thickening, as indicated from optical
observation, is due to cluster breakage and an increase in the
effective volume fraction of the system. The negative normal stress
observed at rest relative to the flowing state appears to originate
during the rapid quench from the tendency of the particles to
collapse into dense clusters on cessation of shear. Accurate
measurements of the normal stress transients as well as the normal
stress dynamics in the quiescent state are the subjects of
continuing work.

The deformation of quenched shear thickened gels by steady flow
($0.1s^{-1} \lesssim \dot\gamma \lesssim 10s^{-1}$) or large
amplitude oscillatory shear ($\omega = 1$ rad/s, $\gamma\geqslant
50\%$) results in the rapid emergence of highly anisotropic
vorticity aligned structures. The transition from shear thickened
3-dimensional gel to an assembly of discrete cylindrical flocs in
rolling flow for a sample of 2 wt.\% carbon black, using a parallel
plate geometry, is shown in Figure \ref{emergence_of_structures}
(see supplemental data for video). In contrast to the structures
observed in the steady state response at low shear rates, these
transient flocs were much more sharply defined and with aspect
ratios exceeding $10^{2}$, spanned macroscopic portions of the shear
cell.

\begin{figure}[h]
\begin{center}
\includegraphics[width=70mm, scale=1]{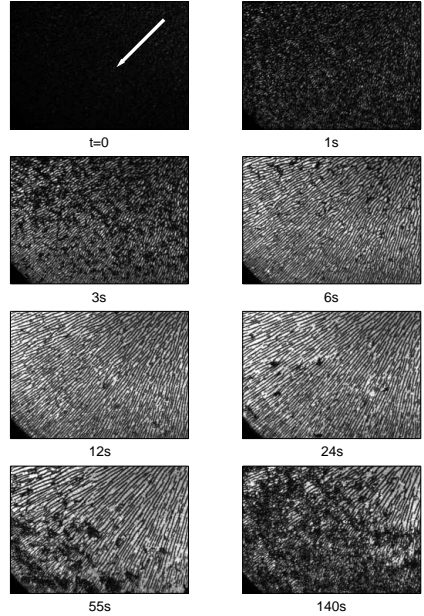}
\caption{Carbon black dispersions, $\phi = 2\, wt. \%$, shear
thickened by strong flow in the secondary thickening regime. Samples
are quenched to zero-shear rate and then subsequently deformed,
starting at $t = 0$ with a shear rate of $\dot\gamma=10s^{-1}$.
Parallel-plate geometry with gap-size $d=50\mu m$. Field of view is
1.35  cm in width. Vorticity direction is along the bottom-left to
top-right diagonal as given by the arrow in the first image.}
\label{emergence_of_structures}
\end{center}
\end{figure}

\begin{figure}[h]
\begin{center}
(a) \includegraphics[width=70mm, scale=1]{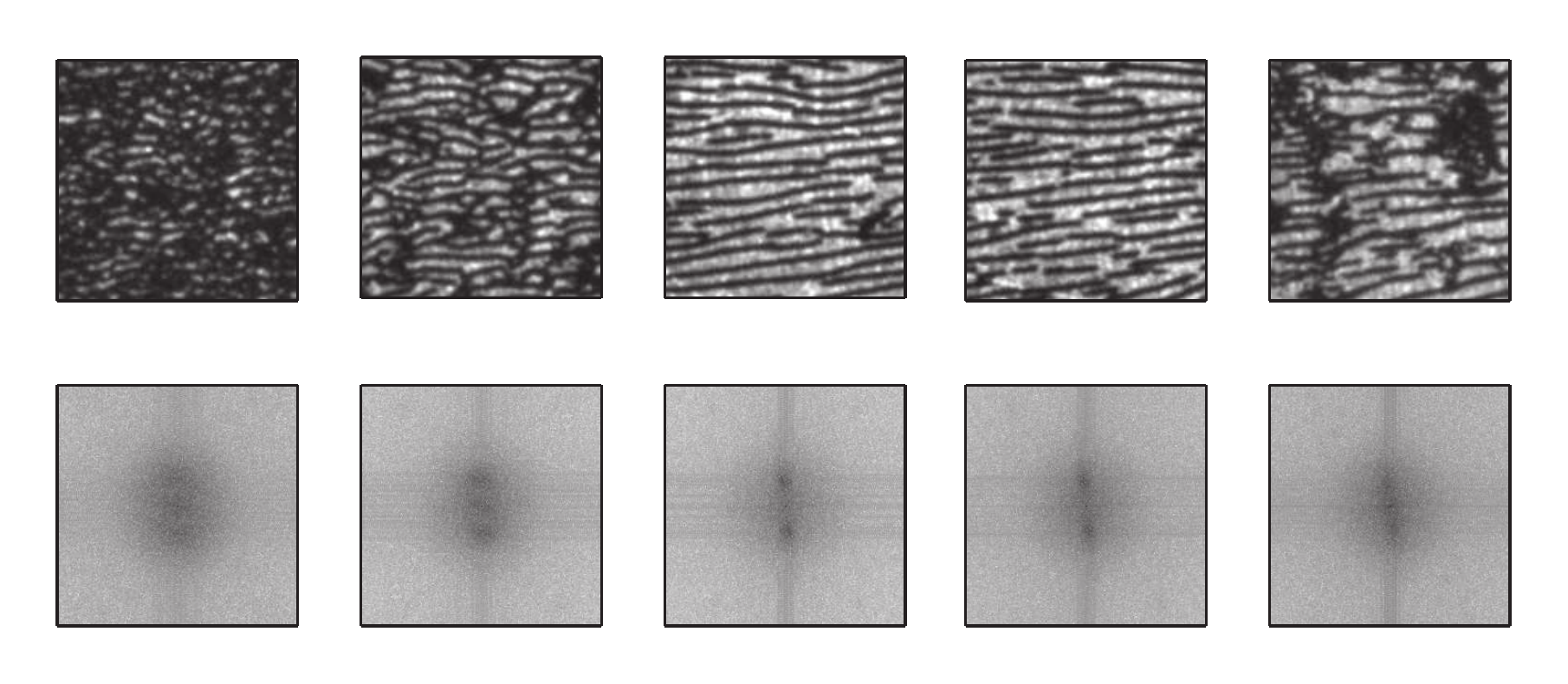}\\
(b) \includegraphics[width=70mm,
scale=1]{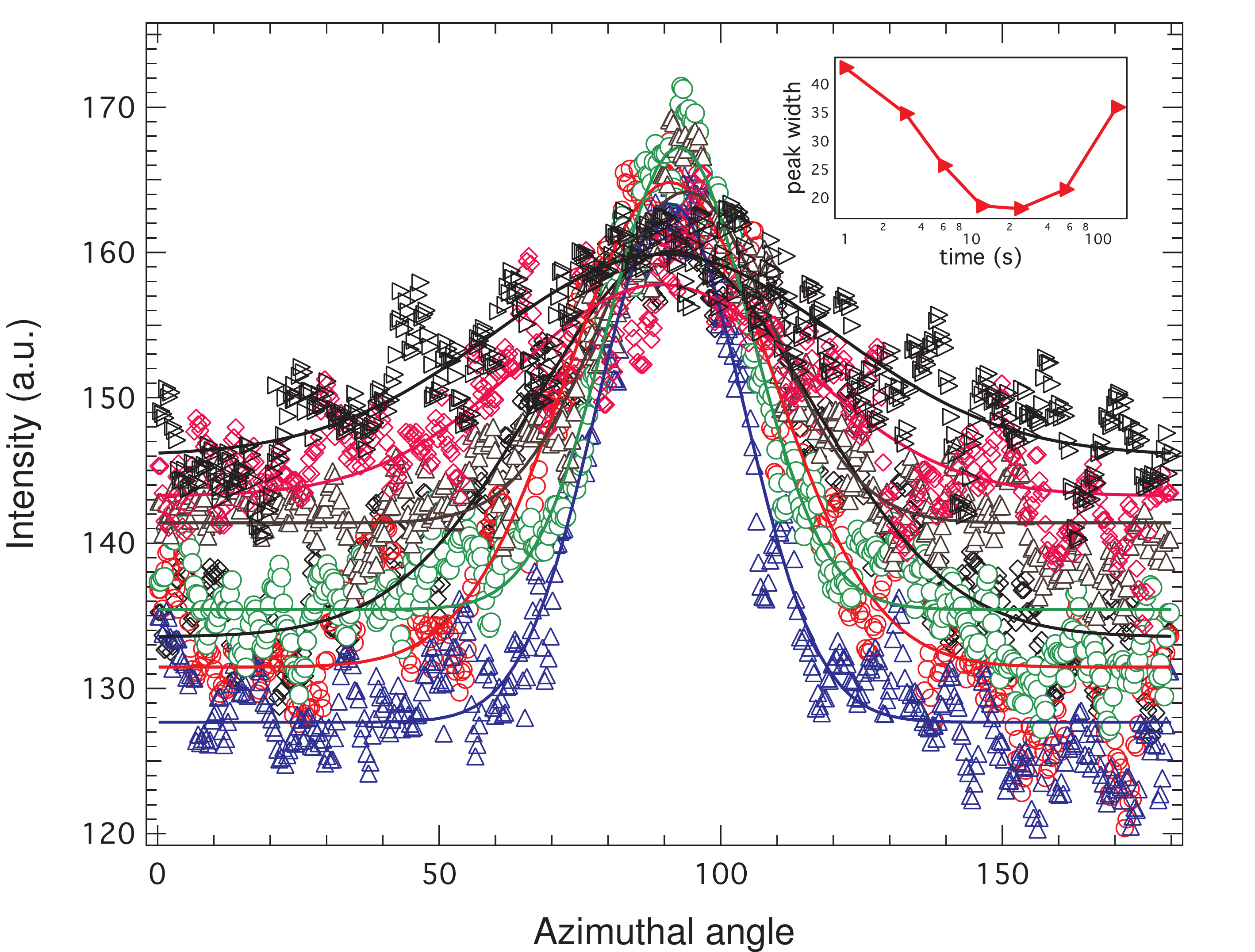} \caption{(a) 3 mm sections
of optical micrographs over time. Gap = 50$\mu m$. From left to
right: $t=1, 3, 12, 24$ and $55s$ (b) Azimuthal spread of intensity
from FFTs of images in (a). Solid lines are Gaussian fits to the
data. Inset: Peak widths vs. log time} \label{ffts_azimuthal_data}
\end{center}
\end{figure}

Fourier transformations of the images provide a means of following
the formation and alignment of the vorticity aligned flocs. We
inspect a small section of the sample, 3 mm on a side and 2.5 mm
from the tool edge, to limit the distortion due to the circular
geometry. We are unable to resolve the rapid changes in the
characteristic length scale of the system due to the low optical
resolution of the camera as well as the short timescale at which a
stable structure emerges, Figure \ref{ffts_azimuthal_data}. We can,
however, monitor the development of alignment in the system. We
extract the azimuthal intensity dependence of the FFTs at the
spatial frequency corresponding to the periodicity of the flocs and
model them with a Gaussian. As shown in Figure
\ref{ffts_azimuthal_data}, alignment is rapid and peaks around 20s,
$\gamma=200$, after which it quickly decreases as the flocs break
up.

\begin{figure}[h]
\begin{center}
\includegraphics[width=70mm, scale=1]{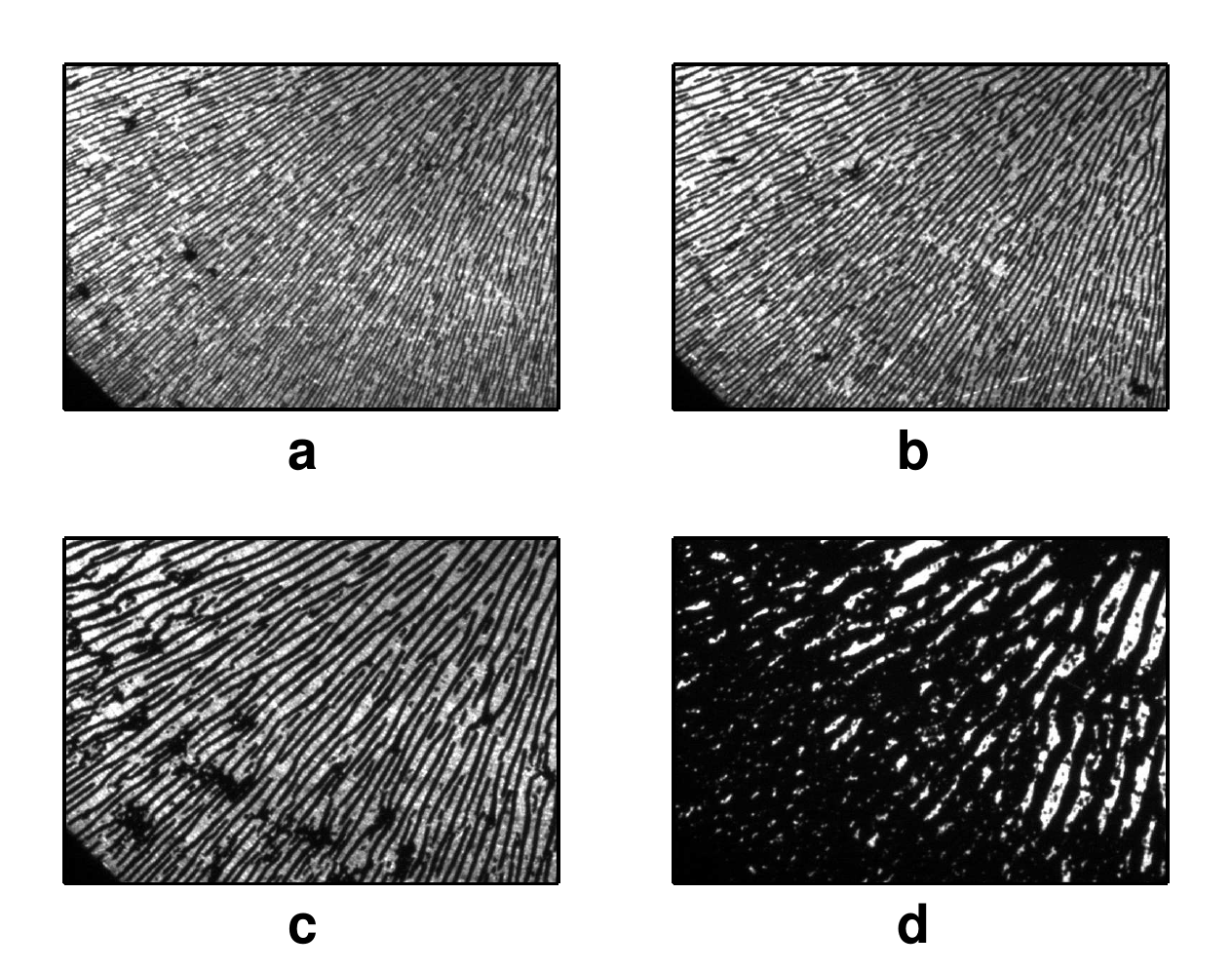}
\caption{Microstructure of 2 wt. \% carbon black dispersions at
$\approx 25s$, $\dot\gamma=10s^{-1}$ after shear thickening flow in
40mm plate-plate geometry with various gaps, $d$. Field of view is
1.35cm in width. (a) $d = 25\mu m$ (b) $d = 50\mu m$ (c) $d = 100\mu
m$ (d) $d = 250\mu m$} \label{gap_dependent_structures}
\end{center}
\end{figure}

Using the parallel-plate geometry, we examined the dependence of the
floc geometry on physical confinement.  The width of the rolling
log-like structures measured at a distance of 2 mm from the tool
edge, $w_{log}$, was moderately larger than the gap size, $d$, and
varied in direct proportion with the gap, Figure
\ref{gap_dependent_structures}. The accuracy in determining the log
width is limited by the $20 \mu m$ per pixel resolution at which
data was taken, but in general, $w_{log}\approx 1.4d$. At gaps
larger than 750 $\mu m$, the structures could no longer be clearly
discerned. From the images, it can be seen that smaller gaps promote
sharper interfaces between the structures and the surrounding fluid.
The width of the structures increases slightly with decreasing shear
rate in moving from the outside of the parallel-plates towards the
center. Increasing gap size results in greater areal coverage of the
parallel-plates as the periodicity of the logs does not increase
commensurate with their width. This implies that the flocs become
less dense internally with increasing thickness. At long times, the
dissolution of the structure proceeds from the outside edge of the
tool, where the shear rate is highest. In observations employing a
low angle cone ($1^{\circ}$) where the shear rate is now constant
across the tool, dissolution proceeds from the outside edge where
the gap is largest, underscoring the importance of confinement in
stabilizing the flocs.

\begin{figure}[h]
\begin{center}
(a) \includegraphics[width=70mm,
scale=1]{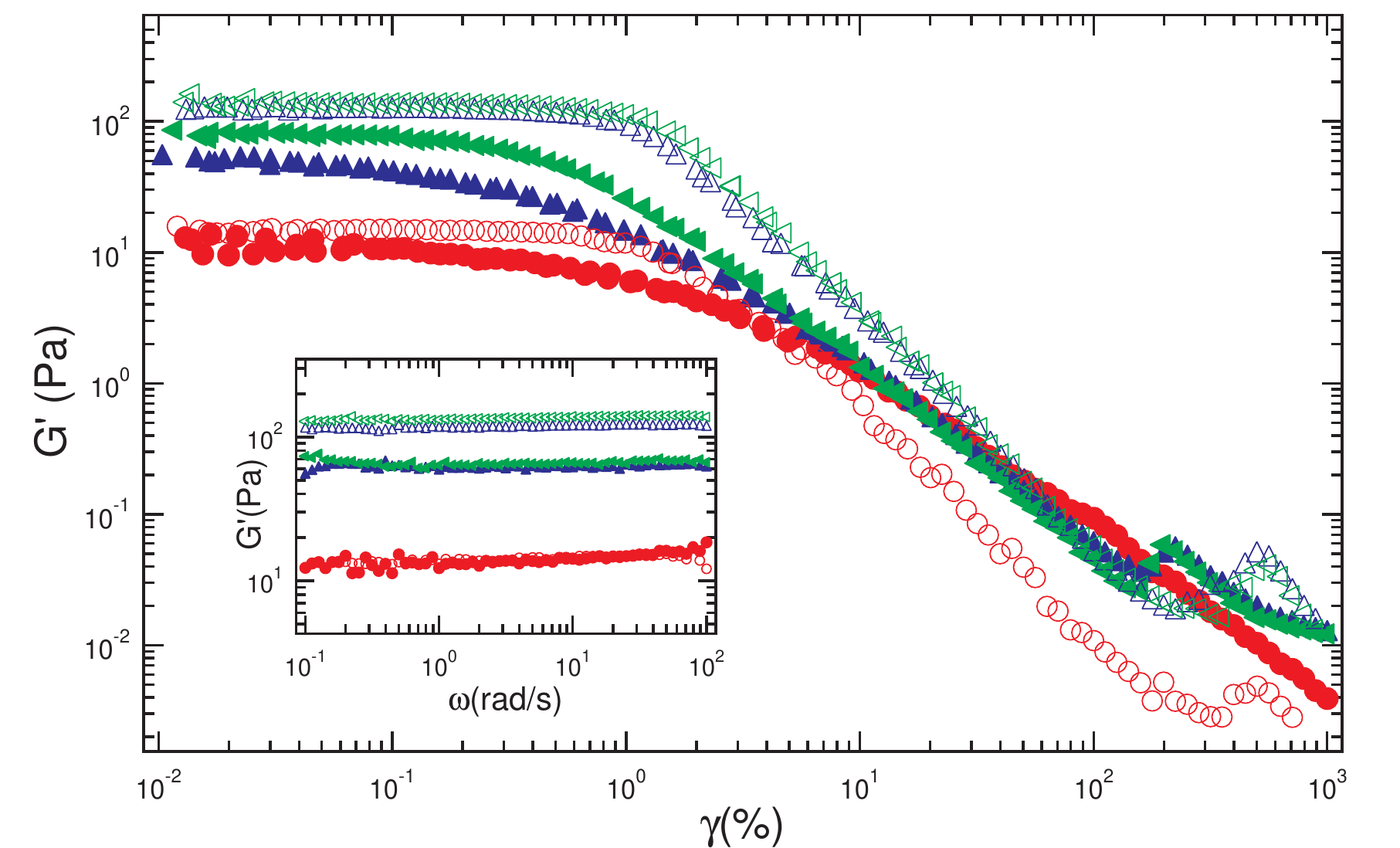}\\
(b)\includegraphics[width=70mm,
scale=1]{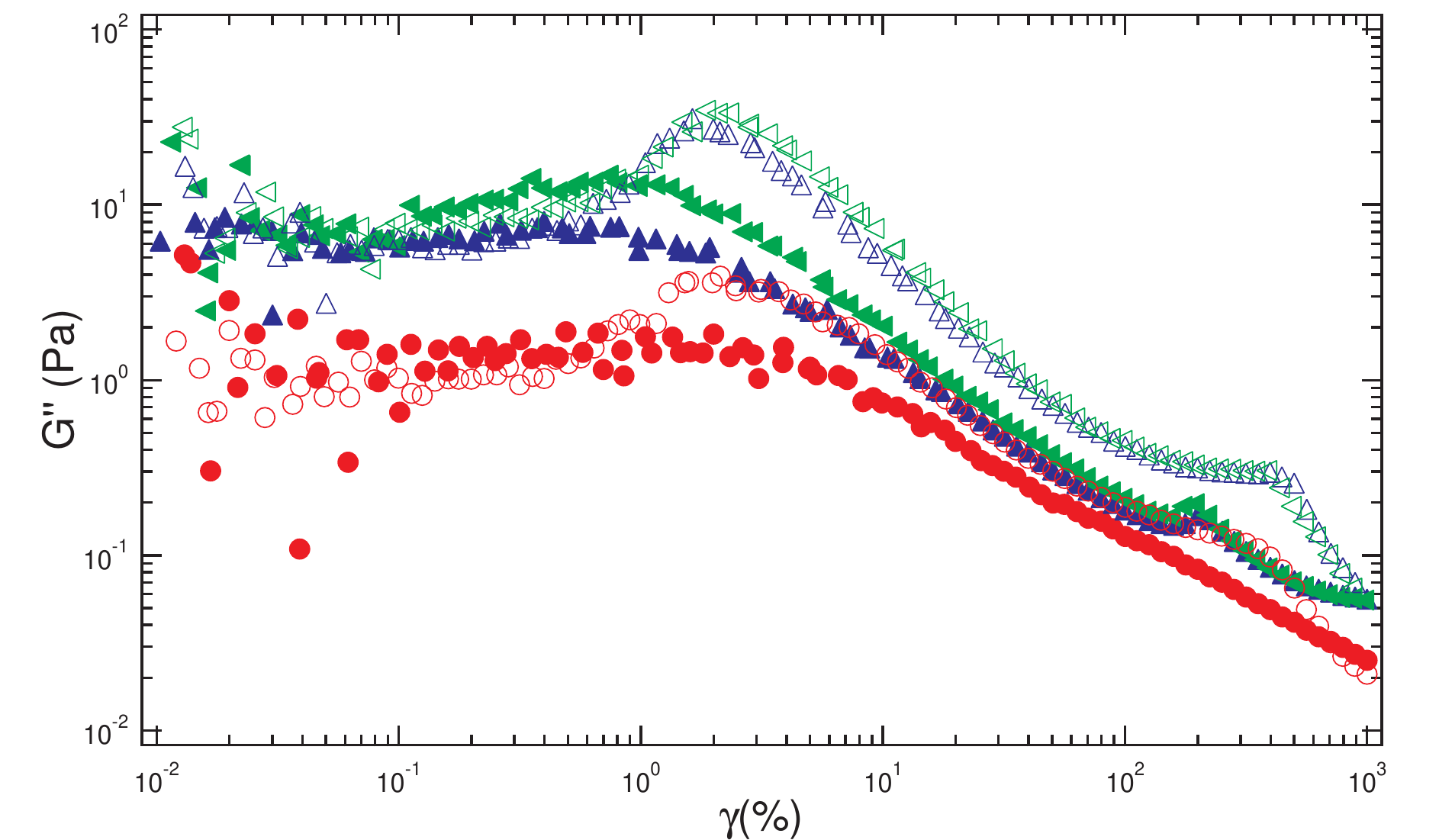} \caption{Strain dependent
moduli of shear thickened gels (open symbols) and log-like flocs
(filled symbols) for carbon black dispersions of various
compositions. Strain data taken at $\omega$ = 1 rad/s. and frequency
data at $\gamma$ = 0.05\%. $\circ$: 2 wt.\%; $\bigtriangleup$: 3
wt.\%; $\triangleleft$: 4 wt.\%.(a) Storage modulus. Inset:
Frequency dependent data. (b) Loss modulus.}
\label{freq_strain_dep_modulus}
\end{center}
\end{figure}

The frequency and strain dependence of the elastic modulus of the
log-like flocs differ from those of the shear thickened gel. Here,
the vorticity-aligned structures were produced by stopping shear
flow at $t=20s$ in a 40 mm, $1^{\circ}$ cone-plate cell with an edge
gap of 350 $\mu m$. Both the gel and the aligned structures show
strong elastic responses, with nearly frequency-independent moduli.
The yield strain, however, is markedly lower for the structures than
for the gel, Figure \ref{freq_strain_dep_modulus}a. Yielding of the
vorticity-aligned structures proceeds without the pronounced upturn
in the loss modulus, $G''$ that typically signifies yielding in the
gel, and the crossover between $G'$ and $G''$ occurs at higher
strains than is typical for the gel. The existence of a yield strain
implies that displacement and deformation in the system is not
accommodated solely by the rolling motion of the flocs.

Vorticity aligned structures have been observed in a variety of
complex fluids such as thixotropic clay gels
\cite{COO:Pignon_PRL1997}, nanotube suspensions
\cite{COO:Hobbie_PRE2004} and attractive emulsion droplets
\cite{COO:Pasquali_PRL2004}, associated in each case with negative
normal stress differences. In our system, the carbon black
dispersions appear to display small negative normal stresses over
all shear rates before increasing at the shear thickening transition
and decreasing again beyond the peak of the transition. No distinct
transition in normal stress was observed under steady flow at low
shear rates, though this is subject to the difficulty involved in
accurate measurements of small changes in these stresses due to
instrument resolution. Measuring the absolute value of the normal
stress is further complicated by the suction of the tool surface by
the sample, i.e. in such systems it is not clear what constitutes
absolute zero. Rather, significance can only be ascribed to
differences in the normal force signal with respect to some initial
condition.

The rapid formation and the high degree of coherence of the
structures formed in the transient flow response of the gel suggest
strongly that relaxation of quenched normal stresses play a
significant role. The dissipation of the normal stress with
continued flow precludes long term stability of the structures and
as a result they exist only transiently. Our results are similar to
observations of an elastic instability associated with flow-induced
clustering in semi-dilute non-Brownian nanotubes
\cite{COO:Hobbie_PRL2004}. Here, the transient rheological response
of a flowing suspension of nanotubes subjected to a shear-rate
quench is underlined by the formation of vorticity-aligned
aggregates. The existence of these aggregates can be parameterized
by confinement and shear rate. In our system, however, we also
observe these structures as a response to steady flow at low shear
rates. By stopping the flow of shear-thickened systems during the
transient response, we have been able to access measurements of the
frequency and strain dependence of the shear modulus of these
structures. The transition from a three-dimensionally connected
network to a two-dimensional system of highly anisotropic structures
and the dissolution of these structures thereafter shows interesting
features. Locally, one can discern changes in topology as the flocs
make and break connections with neighboring structures, and track
changes in the order parameter as shear finally dissolves the system
into non-aligned clusters. The instrument used in this study did not
permit the consistent application of arbitrarily small deformations.
As a result, we were unable to conduct measurements at the low
strain rates that would have permitted better resolution of the
dynamics of structure formation and dissolution in our samples.
Future work will benefit from a more capable instrument that also
permits measurements of normal forces and will pursue more detailed
studies of the effect of confinement and composition on the
mechanics of these structures.
\\
\\
The authors gratefully acknowledge C. Kim and H. Wyss for very
helpful discussions, and the Infineum Corporation for funding.







\bibliography{shear_thickening_vorticity_structures}

\end{document}